\documentstyle[twoside,fleqn,espcrc2not,epsf]{article}

\newcommand{\newc}{\newcommand}
\newc{\mtop}{m_t}
\newc{\mz}{m_Z}
\newc{\mx}{m_X}
\newc{\eg}{{\it e.g.}}
\newc{\ie}{{\it i.e.}}
\newc{\gev}{\,{\rm GeV}}
\newc{\tev}{\,{\rm TeV}}
\newc{\etal}{{\sl et al}}

\title{Leptophobic $U(1)$'s and $R_b$, $R_c$ at LEP}

\author{Chris Kolda\address{School of Natural Sciences, 
	Institute for Advanced Study,
	Princeton, NJ 08540, USA}
	\thanks{Talk given at the IV International Conference on Supersymmetry
	(SUSY-96), College Park, MD, May 29 -- June 1, 1996.
	Research supported in part by Dept.\ of Energy contract
	\#DE-FG02-90ER40542 and by the Monell Foundation.}
	}
       
\begin{document}

\begin{abstract}
In the context of explaining the experimental deviations in $R_b$ and $R_c$
from their Standard Model predictions, a new type of $U(1)$ interaction is
proposed which couples only to quarks. Special attention will be paid to
the supersymmetric 
$\eta$-model coming from $E_6$ which, due to kinetic mixing effects,
may play the role of the leptophobic $U(1)$. This talk summarizes
work done with K.S.~Babu and J.~March-Russell in Ref.~\cite{bkm}.

\end{abstract}

\maketitle

\section{Introduction}
The central lesson which one begins to draw after any preliminary examination
of the data coming from the four experiments at LEP is that the Standard
Model (SM) is in very good shape. In particular, some of the leptonic
measurements at LEP are showing agreement with the SM to nearly 1 part
in 1000, a level of accuracy unusual in high energy experiments. Nonetheless
there are two, by now well known, nagging discrepancies between the LEP data
and the SM, both in the $Z^0$ hadronic decay branching fractions. Defining
$R_q\equiv \Gamma(Z\to q\bar q)/\Gamma(Z\to{\rm hadrons})$, one 
has~\cite{moriond}:
\begin{eqnarray}
R_b&=&\left\lbrace\begin{array}{c}0.2215\pm0.0017\quad{\rm (LEP)}
\\ 0.2152\pm0.0005\quad{\rm (SM)}
\end{array}\right. \\
R_c&=&\left\lbrace\begin{array}{c}0.1596\pm0.0070\quad{\rm (LEP)}
\\ 0.1714\pm0.0001\quad{\rm (SM)}
\end{array}\right.
\end{eqnarray}
where the theoretical uncertainties are dominated by the uncertainty in
the top quark mass, $\mtop=(176\pm 13)\gev$. There is thus a $3.7\sigma$
excess in $R_b$ and a $1.7\sigma$ deficit in $R_c$.

There are a number of plausible solutions to the current disagreement, not
least of which is simple experimental error. $R_b$ alone can be corrected
(at least partially) by low-energy supersymmetry (SUSY)\cite{susyrb}, but
this has no bearing on $R_c$. Solutions have also been proposed in which
new fermions mix with the SM fermion spectrum. The
most attractive of these solutions proposes that the ``top'' quark observed
at FNAL is actually a new fourth generation $t'$-quark, while the real
$t$-quark (defined to be the $SU(2)$ partner of the $b$-quark) is hiding
near the $Z$, $\mtop\approx\mz$~\cite{chw}; this solution again does not
affect $R_c$.

I propose instead an additional $U(1)_X$ interaction whose corresponding
gauge boson, $X$, mixes with the usual $Z$, thereby changing the
interactions of the $Z$ with the SM fermions. Specifically I will build
realistic models in which the $Z$-$X$ mixing can resolve both the $R_b$
and $R_c$ problems without upsetting the other LEP observables; and in
particular I will show that there exists a model coming from $E_6$ which is
phenomenologically acceptable.

\section{Constraints on $U(1)_X$}

Any realistic model of new physics which attempts to resolve the $R_b$ and
$R_c$ anomalies must confront two difficulties. The first difficulty arises
if one wants to assign interactions to the $b$ and $c$-quarks which are not
generation-independent, \ie, change the predictions for only $R_b$ and $R_c$.
 In that case one would generically expect large
flavor-changing neutral currents (FCNC's) through violation of the GIM 
mechanism, particularly in the $D^0$-$\bar D^0$ system.
The solution may lie within the simple observations that, for
$\Delta R_q\equiv R_q^{LEP}-R_q^{SM}$, 
$3\Delta R_b+2\Delta R_c=-0.0047\pm0.0134$, consistent with zero, and that the
total hadronic width at LEP agrees well with the SM. 
Such a pattern of shifts is indicative of a universally-coupled, 
generation-independent $X$-boson:
\begin{eqnarray}
\Gamma_{u,c}&=&\Gamma_{u,c}^{SM}+\Delta\Gamma_c \\
\Gamma_{d,s,b}&=&\Gamma_{d,s,b}^{SM}+\Delta\Gamma_b.
\end{eqnarray}
By having generation-independent couplings, one does not violate the GIM
mechanism and no new large FCNC's are generated.
We will therefore demand this generation-independence as the first of four
principles we will require of any model we build. 

The second difficulty is that the leptonic data at LEP is all very precisely
measured, and in strong agreement with the SM. If the leptons are
charged under $U(1)_X$ then there should appear discrepancies in their
partial widths of roughly the same size as those in the quark sector, an
effect that is clearly not observed. Thus we must demand that the
$U(1)_X$ charges of the leptons be zero (or very small), leading us to name
such interactions {\sl leptophobic}. This leptophobia will be our second
principle for model-building.

Using a leptophobic $U(1)$ to explain the $R_b$ and $R_c$ anomalies was shown
to be phenomenologically viable in Ref.~\cite{hadrophilic}. Though the authors
did not build specific anomaly-free gauge models, there were able to show that
the $Z$-$X$ mixing arising in such a model could in principle explain the 
LEP data. In contrast to~\cite{bkm}, the authors of~\cite{hadrophilic}
also attempted to explain a possible anomaly in the CDF jet cross-section
at high $p_T$. This added requirement led them to consider models which are
very different than those considered here and in~\cite{bkm}: the ``models''
of~\cite{hadrophilic}\ have heavy ($\sim 1\tev$) $X$-bosons and are strongly
coupled, while the models I am considering here have light ($\sim\mz$) 
$X$-bosons and are weakly coupled.

We will impose two further constraints on the models we build, though these
are slightly more prejudical. We will require that all new matter which
is added to the model (\eg, for anomaly cancellation) be vector-like
under the SM gauge groups. This means that the new matter will all receive
masses at the scale of the $U(1)_X$ breaking, which is presumably above
the weak scale. We also require that the SM gauge couplings unify near
$10^{16}\gev$, a behavior expected in grand unified (GUT) and string 
theories and observed
in the minimal SUSY model (MSSM). This will further restrict our new
matter to come in complete multiplets of an ersatz $SU(5)$ (it is ersatz
because the $SU(5)$ need not commute with $U(1)_X$).

Under the conditions laid out above, we built models which fall into
two broad classes~\cite{bkm}. In the first, there are models with extra matter
whose charges have been chosen specifically to cancel the
$G_{SM}\times U(1)_X$ anomalies. Such models could in principle derive from
a string theory, but they do not have the attraction of coming from a
simple GUT group. The second class of models are models
which do come from simple GUT's. That such a model even exists seems at first
improbable since GUT's tend to place quarks and leptons in
common representations, giving charges to quarks and leptons alike
under the GUT subgroups and ruining leptophobia. However I will explain in
the remaining part of this talk just how, in one example, one can indeed
get a leptophobic $U(1)_X$ from a common GUT group, namely $E_6$. But
to do so I must return to the basics of how two abelian gauge bosons can
mix with one another.

\section{U(1) Mixing in an Extended SM}

The most general Lagrangian for the gauge sector of a 
$SU(2)_L\times U(1)_Y\times U(1)_X$ theory can be written:
\begin{eqnarray}
{\cal L}&=&-\frac{1}{4}W_{\mu\nu}W^{\mu\nu}-\frac{1}{4}B_{\mu\nu}B^{\mu\nu}
-\frac{1}{4}X_{\mu\nu}X^{\mu\nu}
\nonumber \\
& &+\frac{1}{2}\mz^2Z_\mu Z^\mu+\frac{1}{2}\mx^2 X_\mu X^\mu \nonumber \\
& &-\frac{1}{2}\sin\chi X_{\mu\nu}B^{\mu\nu}+\Delta m^2Z_\mu X^\mu,
\end{eqnarray}
where $W_{\mu\nu}$, $B_{\mu\nu}$ and $X_{\mu\nu}$ are the usual field
strength tensors. Two terms in this Lagrangian are responsible for
mixing the gauge bosons: there is the usual mass mixing $({\cal L}\sim
\Delta m^2 Z_\mu X^\mu)$ which arises when the SM Higgs fields are charged
under the $U(1)_X$, and the kinetic mixing term $({\cal L}\sim\frac{1}{2}
\sin\chi X_{\mu\nu}B^{\mu\nu})$ which is allowed only because for abelian 
groups the field strengths are themselves gauge-invariant.

Taking the Lagrangian to its mass eigenbasis mixes the $Z$ and $X$ fields
into mass eigenstates, $Z_{1,2}$, where we identify $Z_1$ with the $Z^0$
at LEP. 
One can think of going from the original basis to the physical basis as a two 
step process. First, a 
non-orthogonal transformation diagonalizes the gauge kinetic terms, followed
by a rotation which diagonalizes the mass terms. In the
first of the two transformations, one gets the unexpected result that the 
effective charge to which one of the two $U(1)$'s couples is shifted. That 
is, in the interaction part of the Lagrangian one replaces:
\begin{eqnarray}
{\cal L}_{int}&=&\bar\psi_i\gamma_\mu X^\mu g_X x_i\psi_i \nonumber \\
&\longrightarrow & \bar\psi_i\gamma_\mu X^\mu
\left(\frac{g_X x_i}{\cos\chi}-g_Y y_i \tan\chi\right)\psi_i
\end{eqnarray}
where $x_i$ and $y_i$ are the $U(1)_X$ charge and hypercharge
of $\psi_i$.
After the rotation, the mixing angle, $\xi$, will depend both on $\Delta m^2$ 
and $\sin\chi$:
\begin{equation}
\tan 2\xi\propto \Delta m^2+\mz^2\sin^2\theta_W\sin\chi
\end{equation}
and the interaction of $Z_1$ with matter is given by:
\begin{eqnarray}
{\cal L}_{Z_1}&=&\bar\psi_i\gamma_\mu Z_1^\mu\left[\frac{g_Y}{\sin\theta_W}
(T_{3i}-Q_i\sin^2\theta_W)\cos\xi\right. \nonumber \\
& &\left.+\frac{g_X}{\cos\chi}(x_i+\delta\cdot y_i)
\sin\xi\right]\psi_i
\label{eq:lagrangian}
\end{eqnarray}
where $\delta=-(g_Y/g_X)\sin\chi$.

Now, if one were to begin with a simple GUT at some high scale, then at
the scale at which $G_{GUT}\to U(1)_Y\times U(1)_X\times\cdots$, one has at
tree level $\sin\chi=0$. Then where does a non-zero $\sin\chi$ come from?
Consider $U(1)$ mixing induced by loops, with $X_\mu$ and $B_\mu$ on the
external legs and charged matter running in the loop. The polarization
tensor will have the form:
\begin{equation}
\Pi^{\mu\nu}_{XY}=\frac{1}{6\pi^2}x_iy_i\log\left(\frac{m_i}{\mu}\right)
\left[k^\mu k^\nu-k^2g^{\mu\nu}\right]
\end{equation}
for the $i$-th particle in the loop. This operator in turn generates a term
in the Lagrangian $\frac{1}{2}\sin\chi B_{\mu\nu}X^{\mu\nu}$ where we
identify
\begin{equation}
\sin\chi=\frac{1}{6\pi^2}\sum_i x_iy_i\log\left(\frac{m_i}{\mu}\right).
\end{equation}
Clearly if $\sum_i x_iy_i=0$ for degenerate particles then no mixing is 
induced, and this is often approximately true in GUT's. However,
if the sum is non-zero, then one can resum the logarithms in $\sin\chi$ 
from the GUT scale to the weak scale using the RGE's. (The details of the 
RGE's are found in Ref.~\cite{bkm}.)

\section{$E_6$ and the $\eta$-Model}

Having developed the machinery for calculating $U(1)$ kinetic mixing, let us
now try it on a ``realistic'' GUT group, namely $E_6$. I will not try to 
argue here why $E_6$ is particularly worth studying, since I will assume that
most people are aware of its many useful properties. For now I need only
describe $E_6$ in so far as to note that $E_6\supset G_{SM}
\times U(1)_\chi \times U(1)_\psi$, where the last two $U(1)$ factors
are not contained in the SM. We are considering for this study the simpler
final gauge group $G_{SM}\times U(1)_X$, so we identify $U(1)_X$ as an
arbitrary admixture of the two extra $U(1)$'s in $E_6$:
\begin{equation}
X_\mu=\cos\alpha\, \chi_\mu + \sin\alpha\, \psi_\mu.
\end{equation}
It is a simple exercise to show that no such combination of $U(1)_\chi$ and
$U(1)_\psi$ is leptophobic and can thereby be a candidate for explaining
the LEP anomalies. (In fact, searches for the leptonic decays of the
$\chi_\mu$ and $\psi_\mu$ gauge bosons place bounds on their masses above
$\sim 600\gev$.) 

However, let us add a new term to our Lagrangian, which for
the time being will be simply a phenomenological parameter, namely a non-zero
$\delta\sim\sin\chi$ as in Eq.~(\ref{eq:lagrangian}). 
We then fit all the LEP data, including $R_b$ and $R_c$ (see Ref.~\cite{bkm}
for a discussion of the fitting prodecure) to calculate $\chi^2$
likelihoods for the fits at each choice of $(\alpha,\delta)$. The result,
expressed as 95\% confidence levels are:
$\alpha=-0.89\pm0.06$ and $\delta=0.35\pm0.08$ (the SM is excluded at
$99.5\%$ in this fit). What is remarkable is that
the fit has picked out a very particular subgroup of $E_6$ called the 
$\eta$-model in which $\tan\alpha=-\sqrt{5/3}\simeq-0.91$. Among the 
experimentally interesting characteristics of the $\eta$-model is that the
anomalies of $G_{SM}\times U(1)_\eta$
cancel if and only if entire {\bf 27}'s of $E_6$ live at or below the
$U(1)_\eta$ breaking scale, which is presumably near the weak scale. Thus
large numbers of new particles are expected at scales not very far above a
few hundred GeV. And the reason that the $\eta$-model worked is that
the combination $\eta_i+y_i/3$ vanishes for SM leptons; that is, the
$\eta$-model with $\delta=1/3$ is leptophobic!
%%%%%%%%%%%%%%%%%%%%%%%%%%%%%%%%%%%%%%%%%%%%%%%%%%%%%%%%%%%%%%%%%%%
\begin{figure}
\centering
\epsfysize=2.5in
\hspace*{0in}
\epsffile{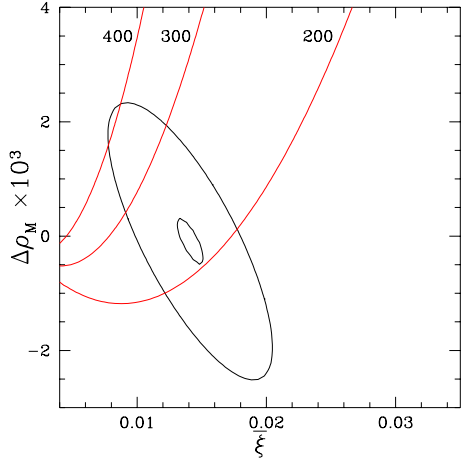}
\caption{The 95\% and 99\% C.L. ellipses for the $\eta$-model with 
$\delta=0.29$. The diagonal lines are contours of the $Z_2$ mass.}
\label{fig}
\end{figure}
%%%%%%%%%%%%%%%%%%%%%%%%%%%%%%%%%%%%%%%%%%%%%%%%%%%%%%%%%%%%%%%%%%%

Getting a realistic spectrum which will generate dynamically $\delta\simeq
1/3$ through the RGE's is not necessarily possible. In fact, it turns out
that there are very few ways in which one can extend the $\eta$-model and
keep it perturbative up to the GUT scale; however, one of these
extensions (see Ref.~\cite{bkm} for details) can generate a value of
$\delta=0.29$, within the 95\% bounds given above, providing a clear example
of dynamical leptophobia. 

In Figure~\ref{fig}, I show the 95\% and 99\% C.L. bounds for the $\delta=0.29$
$\eta$-model in the plane of $(\Delta\rho_M,\bar\xi)$, where $\Delta\rho_M$
is the contribution to the $\rho$-parameter due to $Z$-$X$ mixing and 
$\bar\xi=(g_X\sin\theta_W/g_Y)\xi$ is a rescaled $Z$-$X$ mixing angle.
Cutting across the ellipses of constant $\chi^2$ are contours of the $Z_2$
mass. Rather interestingly, we find that the $\eta$-model predicts rather
light $Z_2$: at 95\%, $200<\mx<240\gev$
(at 99\% C.L. the mass approaches the LEP $Z^0$ mass).

\section{$Z_2$ Signals and Searches}

Traditional methods for searching for and excluding $Z_2$ candidates are
clearly untenable here. One usually produces new gauge bosons through their
interactions with hadrons, but observes their decays
through leptons; this decay channel is simply far too small for a leptophobic
model.
And although the $Z_2$ will decay to jets, those jets are
overwhelmed by QCD backgrounds. To date the best bounds on $Z_2\to jj$ come 
from 
the UA2 Collaboration which was unable to rule out any significant portion
of the mass/coupling range considered here.
It has also been suggested~\cite{bcl}\ that the $Z_2$ could be observed in
associated production with a SM gauge boson, with $Z_2\to b\bar b$; here 
signals and backgrounds are small, but the ratio is close to one.

Finally, it has been realized that low-energy experiments might provide a
strong bound on these extra $Z$ models~\cite{ucb}. In particular, $\nu N$ 
scattering 
and atomic parity experiments can provide an additional window on
$Z$-$X$ mixing. For example, the weak charge of cesium is already
$1.2\sigma$ below the SM value, while the leptophobic models discussed here
all tend to worsen the disagreement. However, it is easy to see that in the
limit $m_{Z_2}\to m_{Z_1}$, 
all $Z_2$ contributions to both low-energy extractions
fall to zero, leaving only the SM prediction. (This is because both
processes require a leptonic coupling at one vertex.) Thus, we are surprised
to learn that not only is a light ($\sim m_{Z_1}$) $Z_2$ allowed, but it 
might even be preferred!

\end{document}